\documentclass[journal,letterpaper]{IEEEtran_nssmic}

\usepackage{cite}      

\usepackage{graphicx}  
\begin{document}
%
\title{WALTA school-network cosmic ray detectors}
%
%
\author{Richard~Gran,~R.~Jeffrey~Wilkes,~Hans-Gerd~Berns,~T.~H.~Burnett%
\thanks{Manuscript received November 14, 2003.  This work was supported in 
part by the U.S. Department of Energy, QuarkNet, and the M.J. Murdock Charitable Trust.}%
\thanks{The authors are at the University of Washington. Contact: gran@phys.washington.edu}%
}
\maketitle

\begin{abstract}
The Washington Area Large-scale Time coincidence Array (WALTA) is placing 
particle detector arrays in secondary schools in the Seattle area to build 
up a large-scale ultra-high energy cosmic ray detector network, one of 
several such projects around the world.  Scintillation counters salvaged 
from the CASA experiment in cooperation with the CROP group at the 
University of Nebraska at Lincoln are refurbished by teachers and students, 
tested, calibrated, and installed in four-fold arrays at high school sites.  
To identify time coincidences, a GPS time synchronization system is employed.
Data are acquired using a custom low-cost data acquisition card.  
Here we will describe 
the logistics of WALTA and show samples of data taken with a 
prototype array at the University of Washington.
\end{abstract}

\begin{keywords}
Cosmic Rays, Air Showers, Secondary Schools.
\end{keywords}

\section{Introduction}
%
%
%
%
\PARstart{P}{article} physics groups around the world are starting outreach 
programs to place cosmic ray detectors at secondary schools in order to study
extensive air showers.
In the Seattle area, secondary schools are spaced nearly
optimally to detect ultra-high energy air showers which
extend for kilometers.  The schools are also spread
over a large area and might detect separate, correlated pairs 
of air showers that are tens of kilometers apart.  
With this method, the schools provide the infrastructure to power the 
detectors and transfer the data to a central location, as well as
enthusiastic volunteers to set up and monitor the detectors. 
In addition to air shower studies, the equipment is also used by 
students and teachers for a variety of elementary experiments
and classroom activities.

In development at the University of Washington 
since 2000, the WALTA project[1][2][3]
now has twenty participating schools in the Seattle area.
Testing of all the pieces of the experimental apparatus is complete.
We will begin taking multi-site coincidence data during the upcoming
school year with the new GPS-based data acquisition card.
 

\section{Elements of the experiment}
Each school is provided with the same set of equipment:  four pieces of
scintillator, four photo-multiplier tubes, a high voltage supply, and a
data acquisition (DAQ) card with a GPS unit.  The school is responsible
for providing boxes to protect the counters from the weather, 
space for counters on
the rooftop of their school or another appropriate location, a computer to
collect the data from the card, and internet access to send the data to 
the university so it can be searched for coincidences. 
The participants may also retrieve data from other sites for 
analysis.

The scintillator, photomultiplier tube (PMT), 
and high voltage module were recovered from the
CASA experiment[5] in the Utah desert by members of WALTA and CROP[4].  
Our teachers and students
refurbish the equipment, polish the edges of the scintillator, and with
our help remount the PMT.  They re-wrap  each unit and test it for light
leaks.

Students characterize the counters by looking at the singles rate
as a function of high voltage and threshold settings and choose appropriate
values.
 
Students and teachers at the schools work either as part of their regular
classroom curriculum or as after-school clubs.  In addition to preparing
the apparatus for measurements of air showers, they perform 
experiments on other sources of radiation, electrical interference, and
tests of muon event rates.  We invite students and teachers to attend
three conferences per year where they can hear about other physics projects
and share their own measurements with their peers.

\subsection{The scintillator counters}
Each piece of scintillator has an area of 0.36 m$^2$ and is 1.27 cm thick.  
The best
choices of high voltage and threshold typically yield a singles rate near
100 Hz, including both muons and the background radioactivity.  
Because our
high voltage modules allow only a single setting, the same for all four
counters at one school, the usual case is that the counter with the lowest 
rate operates at 100 Hz and the others are several times higher.

\subsection{The DAQ card}
\begin{figure}
\centering
\includegraphics[width=3.5in]{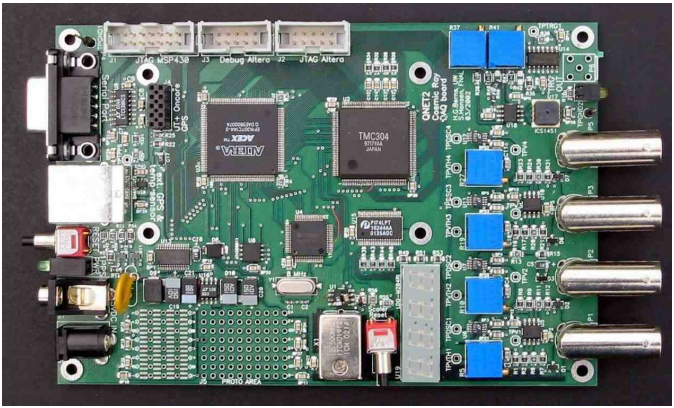}
\caption{QuarkNet DAQ card}
\label{DAQcard}
\end{figure}

The counters are read out by an inexpensive DAQ card[6], shown in Fig. 
\ref{DAQcard} developed with the support of
QuarkNet[7] and in collaboration with engineers at Fermilab and the
University of Nebraska.
The user can select none, 2, 3, or 4-fold coincidence and also select the 
effective gate width from 48 ns to 50 $\mu$s,
based on the fundamental clock cycle of 24 ns. 

When there is a coincidence,
the card reports the rising and falling edges of a discriminator, so it
records when the input pulse goes above and then falls below a threshold
voltage.  The time to digital converter chip 
and its readout circuitry allow for multi-hit events and we
record all edges up to our gate window -- with one limitation -- for each coincidence.   Each edge is recorded with five-bit resolution within this
clock cycle for an actual resolution of 0.75 nanoseconds.
The limitation mentioned above is that this 
chip can store only the first rising and
the first falling edge for each 24 nanosecond clock cycle.  Multiple, closely
spaced edges will be lost.

This design allows us to measure the relative arrival times of the
particles in the air shower to within 0.75 nanoseconds, which is sufficient
to estimate the arrival direction of air showers.  We also plan to
use the time over
threshold to estimate the particle density, and use the multihit capability
to record the width of the shower front for each coincidence.  
With this precision on the card,
we are actually limited by the speed of the PMT's.

\subsection{The GPS unit}
An add-on commercially available GPS unit provides timestamps for each
coincidence.  This is used to match events seen at one school with
particles recorded at other schools.  Based on tests with air shower
data and with pulse generators, two sites can be matched to within
24 nanoseconds in the best case and 350 nanoseconds in the worst case,
depending on the GPS satellites in view.  
The GPS unit is modified 
slightly for our use by the manufacturer to provide timing information
that is not normally available in a GPS navigation device.  We also designed
a custom connector that allows us to use much longer cables than we would 
otherwise.  The method is described
in another paper in these proceedings[8].
Accuracy of tens of nanoseconds is more than sufficient for the 
reconstruction of large air showers and can also be used to search for
coincidences with any similar experiment using GPS.

\section{The prototype arrays}

We have run a prototype site on the rooftop of the physics department at
the University of Washington to test the equipment, make adjustments to
the DAQ card, and to the experimental design.  We have configured the card
to report coincidences that are two-fold or greater 
with an effective gate width of 1.2 microseconds.  
Each of the four channels had a singles rate between 100 and 500 Hz and the
coincidence rate we observed was around 1 Hz, though this varied with
the spatial configuration of the units.  We expect a range of
configurations for each of our participating schools because they will be
limited by the size and shape of their rooftops; typical spacing of modules
in our tests was 6 meters. 

The card was operated with a variety of software:  a basic serial port 
terminal, a LabView interface for Windows, and a command-line interface
on Linux.  We have also developed the essential software to unpack the 
encoded data from the card, test the data for correctness and for 
proper operation of the card, and search the data
for cosmic ray air shower events and multi-site coincidences.

\begin{figure}
\centering
\includegraphics[height=2.5in]{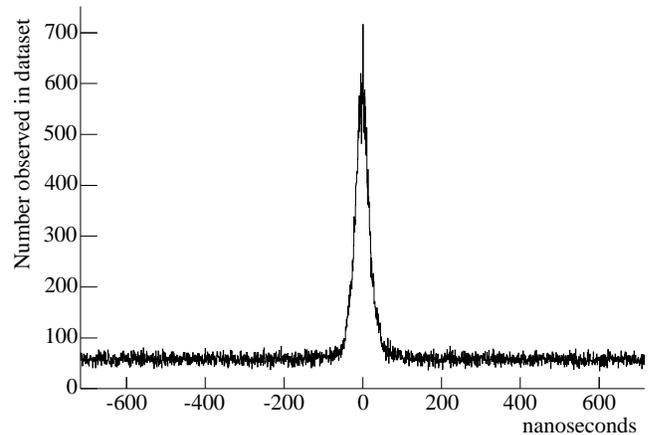}
\caption{Time between rising edges of two-fold coincidences}
\label{twofold}
\end{figure}

\section{Performance}
The distribution of the difference in rising edge times for pairs of
channels for two-fold coincidences is shown in Fig. \ref{twofold}.  
Other pairs of
channels look similar.  The coincidences from air showers are clearly
visible, as is the level of the random background.

\begin{figure}
\centering
\includegraphics[height=2.5in]{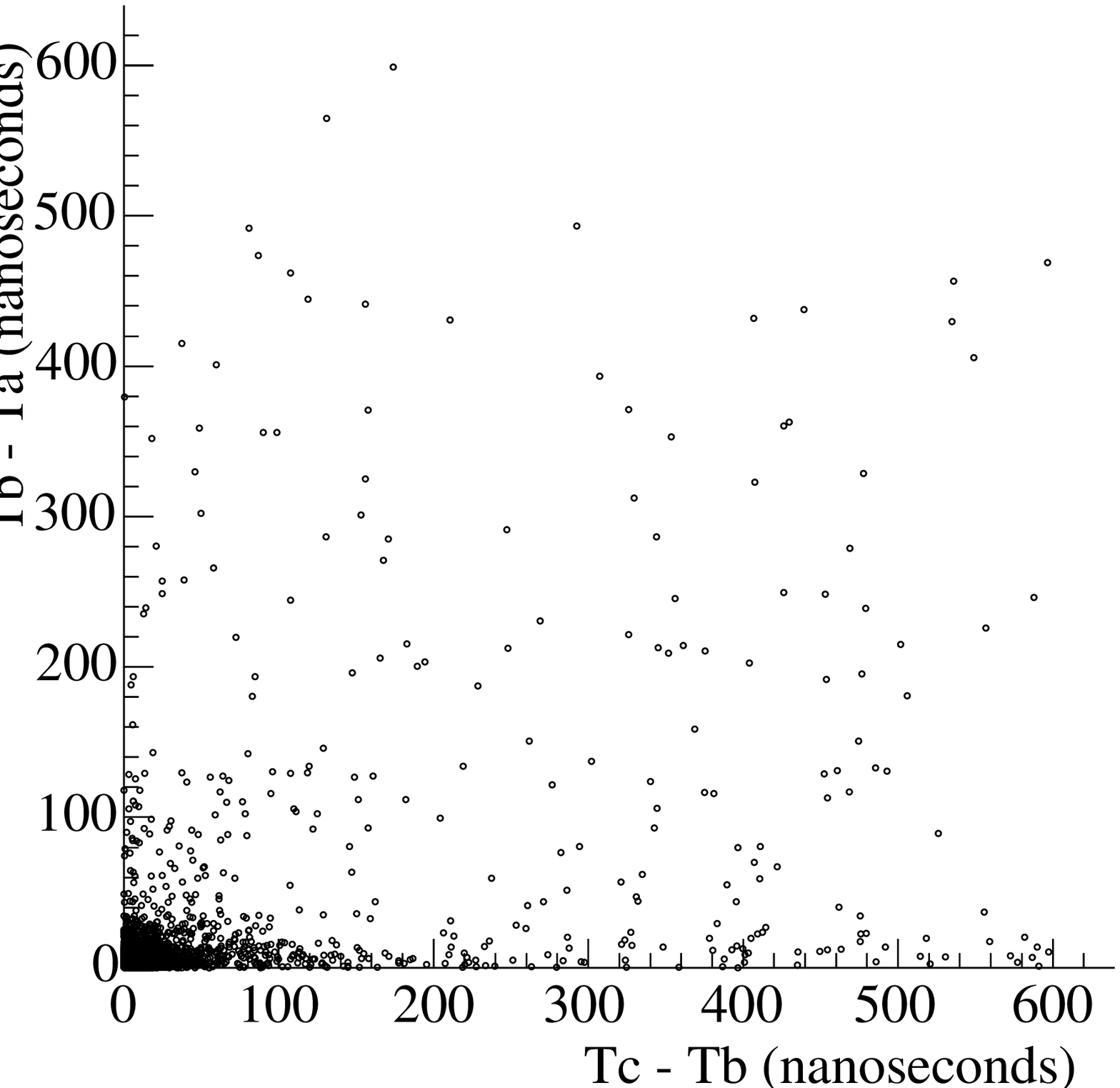}
\caption{Timing of pulses in three-fold coincidences.  
The vertical axis is the time between
the earliest pair of leading edges, horizontal axis is the second pair.}
\label{threefold}
\end{figure}

\begin{figure*}[t]
\centering
\includegraphics[width=5.0in]{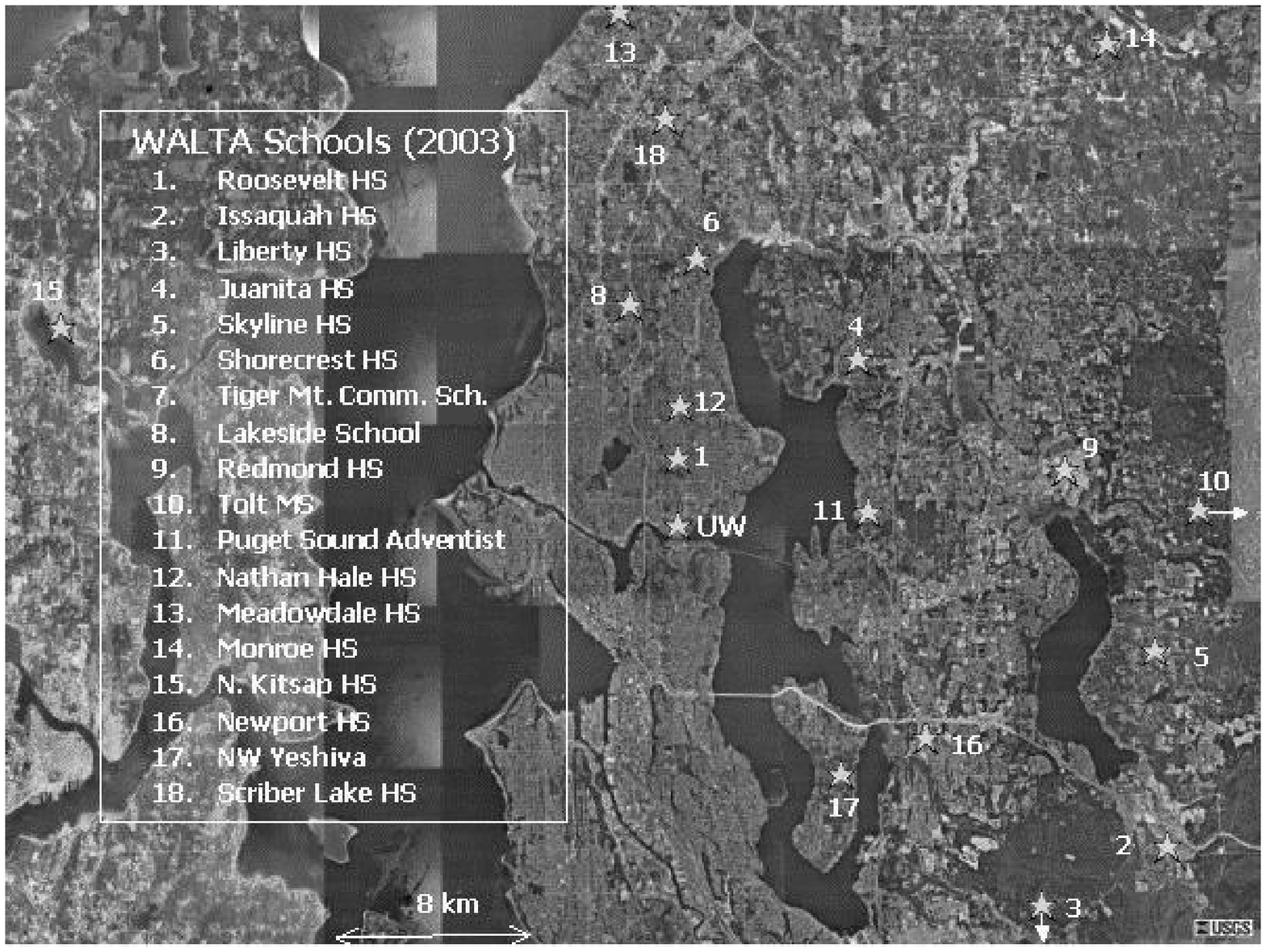}
\caption{The locations of participating WALTA schools}
\label{map.eps}
\end{figure*}

A similar demonstration can be done with three-fold coincidences; because
there is less random background, the air showers with hundred nanosecond structure
are apparent.
We scan through the same data set for these events during offline analysis
and order the leading edge times of the pulses.
There are two time differences for each event, one on each axis
in the scatter plot
in Fig. \ref{threefold}.  As in the previous figure, 90\% of the data
are within 50 ns and are in the lower left corner of the plot.  
The backgrounds in this case are symmetric with respect
to a diagonal of slope one, so the excess most clearly seen along the 
horizontal axis is due to air showers.

%

These data confirm our ability to discriminate both the short and 
the long time structure of air showers.  These structures are
useful for the correlation of ordinary air showers and for ultra-high 
energy cosmic rays, respectively.

We received our first shipment of cards from the production run and
distributed them during our summer workshop for teachers in August 2003.  
At that time we tested the multi-site coincidence capabilities of the GPS
timestamp.  We set four complete arrays on the lawn of the
physics building -- as if they were on school rooftops in the Seattle area.
One of the arrays did not operate correctly; we logged 60 coincidences 
during a two hour run by requiring at least two
of four scintillators in each of the three working arrays.  This test,
along with other tests with pulse generators, 
confirmed the operation of the GPS time-matching system.

\section{Challenges}
Nearly twenty schools have participated in the WALTA program since it 
began, shown in Fig. \ref{map.eps}.  They have prepared their 
scintillator modules and have
been working with NIM electronics for the past couple years.  Nine schools now 
have DAQ cards and are preparing to operate their counters, first in their
classroom, and then shortly afterward put their counters on the rooftops.
The array at the University of Washington has been operating this 
summer and the first school arrays should be operating by the end of 
winter. We expect several multi-site coincidences from ultra-high energy
air showers each month from groups of closely spaced schools.
We will also look for coincidences on city-wide scales and in coordination
with other school-based arrays around the world.
  
In addition to preparing and maintaining the detectors for 
large-area coincidence array, 
our participants have performed many small scale experiments on their own
as part of their curriculum.
We expect they will want to continue to do this, and we will be challenged
to make the data acquisition software used to operate the DAQ card
flexible enough to allow them to do other interesting projects.

After these first sites are operating well we
hope to expand to other schools in the Seattle and King County area.
The area directly south of the University of Washington site is where 
the highest density of schools are, with a spacing of about one kilometer.
Expanding to the rest of King County retains a reasonable density of 
schools while extending the total area covered by this experiment.

The design and testing of the many pieces of the WALTA experiment are 
complete.  The apparatus meets all specifications and has been demonstrated
to detect cosmic ray air showers with very precise particle timing
and also provide accurate two-site event time matching.  The software tools
are available for us and for the students, and they are now being used
at the schools.


%
%


\section*{Acknowledgment}
The authors would like to thank our summer students
Paul Edmon, Jeremy Sandler, Ben Laughlin, and teacher Mark Buchli.



%




\end{document}